\documentclass[dvips]{article}
\usepackage{icrctc07}

\title{Detection of Very-High Energy Gamma-Rays from the BL~Lac Object
PG\,1553+113 with the MAGIC Telescope}

\shorttitle{Detection of VHE Gamma-Rays from PG\,1553+113}

\authors{Robert Wagner,$^{1}$ Daniela Dorner,$^{2}$ Masaaki Hayashida,$^{1}$
Thomas Hengstebeck,$^{3}$\\Daniel Kranich,$^{4}$ Daniel
Mazin,$^{1}$ Diego Tescaro$^{5}$ for the MAGIC collaboration,\\and
Nina Nowak$^{6}$}

\shortauthors{Wagner et al.}

\afiliations{
$^1$ Max-Planck-Institut f\"{u}r Physik, D-80805 M\"{u}nchen, Germany\\
$^2$ Universit\"at W\"urzburg, Am Hubland, D-97074 W\"urzburg, Germany\\
$^3$ Humboldt-Universit\"at zu Berlin, D-12489 Berlin, Germany\\
$^4$ ETH Zurich, CH-8093 Switzerland\\
$^5$ Institut de F\'{\i}sica d'Altes Energies, E-08193 Barcelona, Spain\\
$^6$ Max-Planck-Institut f\"{u}r extraterrestrische Physik,
D-85748 Garching, Germany} \email{robert.wagner@mppmu.mpg.de}

\abstract{The MAGIC telescope has observed very-high energy
gamma-ray emission from the BL~Lac object PG 1553+113 in 2005 and
2006 at an overall significance is $8.8 \sigma$. The light curve
shows no significant flux variations on a daily timescale. The
flux level during 2005 was, however, significantly higher as
compared to 2006. The differential energy spectrum between
$\sim$90 GeV and 500 GeV is well described by a power law with a
spectral index of $-4.2\pm0.3$. The photon energy spectrum and
spectral modeling allow to pose upper limits of $z=0.74$ and
$z=0.56$, respectively, on the yet undetermined redshift of PG
1553+113. Recent VLT observations of this blazar show featureless
spectra in the near-IR, thus no direct redshift could be
determined from these measurements.}

\begin{document}
\maketitle
\section{Introduction}
The active galactic nucleus (AGN) PG~1553+113 was
first reported in the Palomar-Green catalog of UV-bright objects
\cite{green}. Its spectral characteristics are close to those of
X-ray selected BL~Lacs \cite{falomo90}. Despite several attempts,
no emission or absorption lines have been found in the spectrum of
PG 1553+113. Thus only indirect methods can be used to determine
the redshift $z$ \cite{sbarufatti05, sbarufatti06}. Recently the
H.E.S.S. \cite{hess1553} and MAGIC \cite{magic1553} collaborations
have presented $\gamma$-ray signals at the $>4\sigma$ and the
$8.8\sigma$ level, respectively, making this source a confirmed
VHE $\gamma$-ray emitter. Here we present the MAGIC observation
results.
%
%
\section{Observation and data analysis}
PG 1553+113 was observed for 8.9~h in April/May 2005, and for 19~h
from January to April 2006 with the MAGIC telescope
\cite{magictech}, the currently largest Cerenkov telescope with a
low trigger threshold of $\approx$60~GeV. Data taken during
non-optimal weather conditions or affected by hardware problems
were excluded from the analysis. Also, only data taken at small
zenith angles ZA$<30^\circ$ were used. After these selection cuts,
7.0~h and 11.8~h of good data remained for 2005 and 2006,
respectively. Given the mean ZA of $\sim 22^{\circ}$, $\gamma$-ray
events above $\sim 90~$GeV have been used for the physics
analysis. In addition, Off-data were taken on a nearby sky
position with comparable ZA distribution and night sky conditions.
These data were used to determine the background content in the
signal region of the on-data. In total, 14.5~h of Off-data (6.5~h
from 2005 and 8.0~h from 2006) were used for the analysis. Since
the two off-samples were in good agreement we used the combined
data to analyze the individual On-data samples.
The data were analyzed using the standard MAGIC analysis chain
\cite{bretz03,gaug05,wagner05}. The method for discrimination
between hadron-induced and $\gamma$-ray-induced showers is based
on the Random Forest (RF) method \cite{breiman}, which was trained
on Off-data and Monte Carlo (MC) generated
\cite{knapp04,majumdar05} $\gamma$-ray events. The significance of
any excess was calculated according to Eq.~17 in \cite{li83} where
the On/Off ratio $\alpha$ was derived, taking into account the
smaller error from the Off-data fit. The RF method was also used
for the energy estimation of the $\gamma$-ray showers, resulting
in an average energy resolution of 24\% RMS.
Simultaneously with MAGIC, optical observations were performed
with the KVA telescope.

\section{Results}
\begin{figure}[h!]
\begin{center}
\includegraphics[width=.8\linewidth]{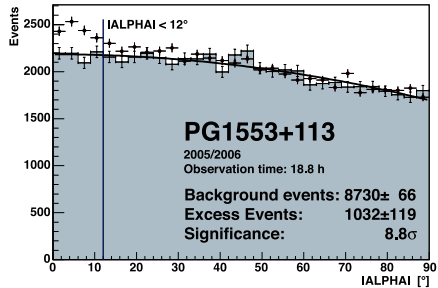}
\end{center}
\vspace*{-0.3cm} \caption{ALPHA plot for the combined 2005 and
2006 PG\,1553+113 data after cuts. The diagram also shows the
distribution of the (normalized) Off-data and a second-order
polynomial describing the off-data. \label{fig:alphaplot}}
\end{figure}

Combining the data from 2005 and 2006, a very clear signal is seen
in the image parameter ALPHA, as shown in
Fig.~\ref{fig:alphaplot}. Defining the signal region as
ALPHA$<12^\circ$ (containing about 90\% of the $\gamma$-ray
events), an excess of 1032 over 8730 background events yields a
total significance of $8.8~\sigma$. The individual results for the
years 2005 and 2006 are listed separately in
Table~\ref{tab:results}.

\begin{table*}
\begin{tabular}{c|c|c|c|c|c|c|c} Year & on time & N$_\mathrm{on}$ &
N$_\mathrm{off}$ & on/off & sigma & $F \left( E > 200 \mathrm{GeV}
\right)$ & photon index
\\
\hline
 2005 & 7.0~h & 3944 & $3501 \pm 26$ & 0.20 &
$6.7\
\sigma$ & $2.0 \pm 0.6_\mathrm{stat} \pm 0.6_\mathrm{sys}$ & $4.31 \pm 0.45$ \\
2006 & 11.8~h & 5815 & $5228 \pm  39$ &  0.30 & $7.0\
\sigma$ & $0.6 \pm 0.2_\mathrm{stat} \pm 0.2_\mathrm{sys}$ & $3.95 \pm 0.23$ \\
2005+2006 & 18.8~h & 9761 & $8730 \pm 66$ & 0.49 & $8.8\ \sigma$ &
$1.0 \pm 0.4_\mathrm{stat} \pm 0.3_\mathrm{sys}$ & $4.21 \pm 0.25$
\end{tabular}
\vspace*{-0.2cm} \caption{Results from the PG 1553+113 analysis as
derived for 2005 and 2006. Integral flux in units of $10^{-11}\
\mathrm{cm^{-2}\ s^{-1}}.$\label{tab:results}}
\end{table*}

\begin{figure}
\begin{center}
\includegraphics[width=.85\linewidth]{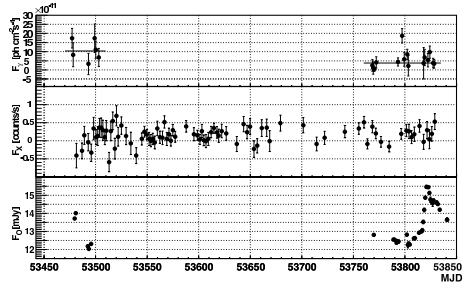}
\end{center}
\vspace*{-0.4cm} \caption{VHE $\gamma$-ray ($120~\mathrm{GeV}$ --
$400~ \mathrm{GeV}$), X-ray ($2~\mathrm{keV}$ --
$10~\mathrm{keV}$) and optical light curve ($R$-band) of PG
1553+113 in 2005 and 2006. The horizontal bars in the top panel
correspond to the average flux during 2005 and 2006, respectively.
The X-ray data were obtained from the All-Sky-Monitor on board the
RXTE satellite. \label{fig:lightcuve}}
\end{figure}

The $\gamma$-ray, X-ray and optical light curve of PG\,1553+113
are shown in Fig.~\ref{fig:lightcuve}. While the optical data show
significant short term variability on the 25\% level, the X-ray
data are compatible with a constant emission (weighted mean of
$0.15 \pm 0.03~\mathrm{counts/s}$). In $\gamma$-rays there is no
evidence for short term variability, but a significant change in
the flux level from 2005 to 2006 is found, given a systematic
error of the analysis on the flux level of about 30\%. The average
integral flux between $120~ \mathrm{GeV}$ and $400~ \mathrm{GeV}$
is given as $F = 10.0 \pm 0.23_\mathrm{stat}$ and $F = 3.7 \pm
0.08_\mathrm{stat}$ (with $F$ given in units of $10^{-11}\
\mathrm{cm^{-2}\ s^{-1}}$) for 2005 and 2006, respectively. On
2006 February 25, prior to the optical flare, a degree of optical
linear polarization of $8.3 \pm 0.2\%$ and a polarization angle of
$139.1^\circ \pm 0.4^\circ$ was measured. Since the host galaxy
cannot be resolved, the optical flux should correspond to the
emission from the AGN core.

The combined 2005 and 2006 differential energy spectrum for PG
1553+113 is shown in Fig.~\ref{fig:spectrum_1}. The integral
fluxes above 200~GeV and the spectral slope coefficients for the
different samples are listed in Tab.~\ref{tab:results}, taking
into account the full instrumental resolution \cite{mizobuchi05}.
The energy spectrum is well described by a pure power law:
\begin{figure}
\begin{center}
\includegraphics[width=.8\linewidth]{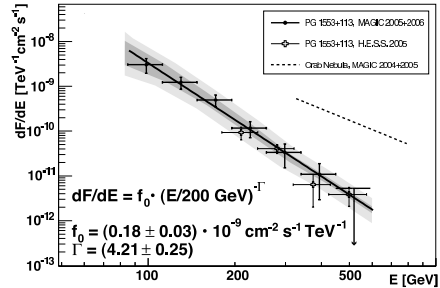}
\end{center} \vspace*{-0.3cm} \caption{Differential energy spectrum
of PG 1553+113 as derived from the combined 2005 and 2006 data.
The MAGIC Crab energy spectrum and the H.E.S.S. PG 1553+113 energy
spectrum have been included for comparison.
\label{fig:spectrum_1}}
\end{figure}
\begin{equation}
\frac{\mathrm{d}N}{\mathrm{d}E}=\left( 1.8 \pm 0.3_\mathrm{stat}
\right) \cdot \left( \frac{E}{200\ \mathrm{GeV}} \right) ^{-4.2
\pm 0.3_\mathrm{stat}}
\end{equation}
\noindent(in units of $10^{-10}\mathrm{cm^{-2}\ s^{-1} \
TeV^{-1}}$, $\chi^2 / \mathrm{d.o.f.} = 1.5 / 4$). The spectral
slopes of the individual years are in good agreement although the
flux level above $200~ \mathrm{GeV}$ is about a factor 3 larger in
2005 as compared to 2006. The estimated systematic error of the
analysis (signal extraction, cut efficiencies etc.) is 25\% (dark
colored band in the figure) and 30\% on the energy scale (light
colored band).

\section{Discussion}
The MAGIC detection confirms the tentative signal seen by H.E.S.S.
at a higher energy \cite{hess1553}. The agreement between the
H.E.S.S. and MAGIC energy spectra is reasonably good. While the
spectral slope is consistent within errors, the absolute flux
above 200~GeV in 2005 is by a factor 4 larger compared to H.E.S.S.
This difference may in part be explained by the systematic errors
of both measurements but also by variations in the flux level of
the source (the observations with H.E.S.S. were commenced after
MAGIC). The observed energy spectrum is steeper than that of any
other known BL~Lac object. This may be an indication of a large
redshift ($z > 0.3$), but can as well be attributed to intrinsic
absorption at the AGN or, more naturally, to an inverse Compton
peak position at lower energies. The spectrum can, however, be
used to derive an upper limit on the source redshift from physical
constrains on the intrinsic photon index ($\Gamma_\mathrm{int} >
1.5$) as discussed in \cite{hess1553}. Using the lower limit on
the evolving EBL density from \cite{kneiske04} we derived a $2
\sigma$ upper limit on the redshift of $z < 0.74$, which agrees on
the limit found in \cite{hess1553}. Other approaches to limit the redshift are given in \cite{bc,fg}.

\begin{figure}
\includegraphics[width=\linewidth]{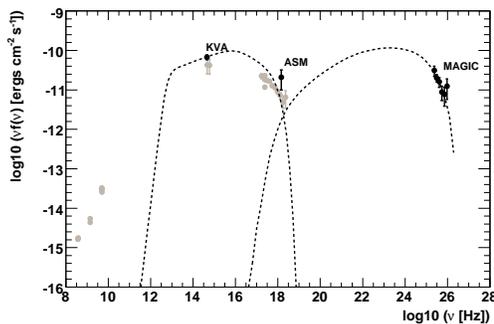}
\vspace*{-1.2cm} \caption{SED of PG 1553+113. The solid lines are
the result of a SSC model fit to the black data points using the
code provided by \cite{krawczynski04}. The gray points comprise
non-simultaneous radio, optical and X-ray data from
\cite{giommi02}.} \label{fig:spectrum_2}
\end{figure}

The spectral energy distribution (SED) of PG 1553+113 together
with the results from a model calculation are shown in
Fig.~\ref{fig:spectrum_2}. The VHE data points correspond to the
intrinsic spectrum of PG 1553+113 as derived for a redshift of $z
= 0.3$. The black points at low energies denote the average
optical and X-ray flux taken at the same time as the MAGIC
observations. Non-simultaneous radio, optical and X-ray data were
taken from \cite{giommi02}. The solid line shows the result of a
model fit to the simultaneously recorded data (black points) using
a homogeneous, one-zone Synchrotron Self-Compton (SSC) model
provided by \cite{krawczynski04}. The $\gamma$-ray, X-ray and
optical data are well described by the model. This is not the case
for the radio data, where intrinsic absorption requires a much
larger emitting volume compared to X-rays and $\gamma$-rays.
Except for a somewhat smaller radius of the emitting region,
identical model parameters as in \cite{costamante02} have been
used: Doppler factor $D=21$, magnetic field strength $B=0.7\
\mathrm{G}$, radius of the emitting region $R=1.16^{+0.62}_{-0.21}
\cdot 10^{16}\ \mathrm{cm}$, electron energy density $\rho_e =
0.11^{+0.18}_{-0.06} \ \mathrm{erg / cm^3}$, slope of the electron
distribution $\alpha_e=-2.6$ for $8.2 < \log \left( E /
\mathrm{eV} \right) < 9.8^{+0.2}_{-0.05}$ and $\alpha_e=-3.6$ for
$9.8^{+0.2}_{-0.05} < \log \left( E / \mathrm{eV} \right)
<10.6^{+1.6}_{-0.0}$. The limits on some of these parameters
indicate the change of the SED model parameters when varying the
assumed redshift from $z=0.2$ up to $z=0.7$ (parameters without
limits were kept constant for all fits). In the case of $z \ge
0.56$ the SED model cannot accurately describe the data and, based
on the obtained $\chi^2$ value for the SED fit, $z>0.56$ is
excluded on the $4.5\sigma$ level.

PG 1553+113 was in a high state in the optical in both years
showing a strong flare at the end of March 2006. The high linear
polarization of the optical emission ($8.3 \pm 0.2 \%$) indicates
that a sizeable fraction of the optical flux is indeed synchrotron
radiation. In $\gamma$-rays only a significant change in the flux
level from 2005 to 2006 is found while there is no evidence for
variability in X-rays. As a result, a possible correlation between
the different energy bands cannot be established. A possible
connection between the $\gamma$-ray detection and the optical high
state can, however, not be excluded. The optical flare without
X-ray or $\gamma$-ray counterpart may still be explained by
external-inverse-Compton models which predict a time lag of the
X-rays and $\gamma$-rays with respect to the optical emission.

\section{SINFONI Near-IR Spectroscopy}
Previous attempts to measure the redshift of PG\,1553+113 were
based on optical spectra. As in near-IR the nucleus outshines the
galaxy's spectral features less than in the optical, archival ESO
data taken with the integral-field spectrograph SINFONI
\cite{Eisenhauer-sinfoni} at the VLT\footnote{Based on
observations at the European Southern Observatory VLT
(276.B-5036(A)), obtained from the ESO/ST-ECF Science Archive
Facility} were analyzed. The data were obtained without adaptive
optics using the largest field of view ($8\times8$ arcsec) and the
H+K grating during four nights in March 2006.
We used 135~min of data obtained under good seeing conditions
(PSF\,$\approx0.4$) and carefully removed the strong sky
background using \cite{Davies-07}.
No host galaxy was detected in the images and the spectra are
featureless, thus no redshift can be determined.

\section*{Acknowledgments}
We thank the IAC for the excellent working conditions in La Palma
and the RXTE team for providing the ASM X-ray data. We further
acknowledge the support of BMBF, MPG, INFN, and CICYT. This work
was also supported by ETH Research Grant TH-34/04-3 and by Polish
Grant MNiI 1P03D01028.

\end{document}